\documentclass[twocolumn,pre]{revtex4}  
\usepackage{graphicx}
\usepackage{dcolumn}
\usepackage{bm}
\usepackage{amsxtra}
\usepackage{latexsym}
\usepackage{subfigure}

\begin{document}
  
\title{Scaling behavior of the disordered contact process}  

\author{S.~V.~ Fallert}
 \email{sf287@cam.ac.uk}  
\affiliation{Department of Chemistry, University of Cambridge,  
             Cambridge, UK }

\author{S.~N.~Taraskin}  
\affiliation{St. Catharine's College and Department of Chemistry, University of Cambridge,  
             Cambridge, UK}  

\date{\today}

\begin{abstract} 
The one-dimensional contact process with weak to intermediate quenched
disorder in its transmission rates is investigated via
quasi-stationary Monte Carlo simulation.
We address the contested questions of both the nature of dynamical scaling,
conventional or activated, as well as of universality of critical
exponents by employing a scaling analysis of the distribution of
lifetimes and the quasi-stationary density of infection. 
We find activated scaling to be the appropriate description for all
disorder strengths considered.
Critical exponents are disorder dependent and approach the values
expected for the limit of strong disorder as predicted by
strong-disorder renormalization group analysis of the process.
However, even for the strongest disorder under consideration no
strong-disorder exponents are found.
\end{abstract} 

\maketitle  
  

The critical behavior of systems with quenched randomness has been
the subject of interest for a long time.
Over the past decades, their investigation has
revealed rich behavior including the existence of new phases
\cite{Griffiths_69}, novel fixed points \cite{igloi_05}, and 
unconventional scaling properties \cite{fisher_87}.

Recently, attention has turned towards the influence of disorder on 
stochastic many-particle systems with a phase transition into an
absorbing state owing to their relevance in physics, chemistry and 
biology \cite{Hinrichsen_00:review}.
In particular, the contact process (CP) \cite{harris_74}, a paradigmatic
model for the stochastic spreading of an infectious disease, has been
investigated as a representative for the prominent universality class of
directed percolation (DP).
Interest in the influence of disorder on this process was sparked by
the surprising lack of experimental observation of DP behavior in real
systems, for which disorder may be responsible \cite{Hinrichsen_00}.
With a recent study presenting convincing evidence of DP critical
behavior in turbulent liquid crystals \cite{takeuchi_07}, an
understanding of the effects of disorder is more relevant than ever.

Initial Monte Carlo (MC) studies of the disordered CP (DCP)
 found continuously varying dynamical critical exponents assuming
 conventional scaling \cite{noest_85,dickman_96,Cafiero_98}.
Recently, deep insight was gained through a strong-disorder
renormalization group study of the DCP  which revealed an
infinite-randomness fixed point (IRFP) for sufficiently strong
disorder in close analogy to the random transverse-field Ising model
\cite{Hooyberghs_03}.
While this makes the strong-disorder limit of the process well
understood \cite{igloi_05} and predicts new strong-disorder
exponents as well as an unconventional ``activated'' form of dynamical
scaling, the behavior in the weak- and intermediate-disorder regime
remains a subject of
debate \cite{Hooyberghs_04,Vojta_05}.
Initial density-matrix renormalization group (DMRG) studies were not
able to convincingly distinguish the two
alternative dynamical scaling scenarios, conventional or activated, 
and reported
critical exponents continuously changing with disorder strength
\cite{Hooyberghs_04}.
In contrast, a recent MC study reported evidence for
activated scaling with strong-disorder exponents for all disorder
strengths \cite{Vojta_05}.

Static scaling in the DCP was found to be of conventional form
\cite{dickman_98} as predicted by the strong-disorder renormalization
study \cite{Hooyberghs_03}.
However, there exists conflicting evidence as to the universality of
the exponents for weak and intermediate disorder with the literature
reporting both disorder-dependent \cite{dickman_98,Neugebauer_06} and
strong-disorder \cite{Vojta_05} exponents.

In this paper, we aim to address both the question of the type of
dynamical scaling as well as of universality of exponents in the
weak- and intermediate-disorder regime by considering the scaling of
the distribution of lifetimes, $P(\tau)$, of the 1d DCP obtained from
quasi-stationary MC simulation.
This is motivated by the fact that an analysis of the scaling behavior
of entire distributions promises to yield clearer results as compared
to the scaling of means \cite{Young_96,Hooyberghs_04}.
Further, in dynamic single-seed MC simulations employed for the
DCP in the past \cite{dickman_96,Vojta_05}, the question of whether the
long-time limit of the process had been reached was frequently
contested.
In contrast, quasi-stationary simulations offer a clear means of
ensuring this: a true stationary average whose convergence can be
monitored.

In the clean CP (without disorder)  defined on a lattice, sites represent the individuals
of a population which can be in two possible states, susceptible or infected.
An infected site attempts to spread its infection to nearest
neighbors at rate $\lambda$, while recovery is spontaneous at rate
$\epsilon=1$.
In the limit $t \to \infty$ for an infinite system, there exist two distinct
states: an active one where a finite density $\rho$ of infected sites 
remains and a non-active regime in which the system ultimately gets
trapped in an absorbing state with no infected sites remaining.
The system undergoes a continuous phase transition between these two
phases at a critical rate  $\lambda - \lambda_c^0 \equiv \Delta = 0$
with order parameter $\rho$ \cite{Hinrichsen_00:review,Marro_99:book}.

Starting from a fully-infected system, the density of infected sites
initially relaxes while spatial correlations grow towards the size
of the system and temporal correlations decay.
Once the correlation length becomes comparable to this size, the
process enters a quasi-stationary (QS) regime.
This metastable state is characterized by a non-zero time-independent
transition rate to the absorbing state.
Given that no true non-trivial stationary state can exist in a finite
system, in these cases QS averages are commonly used as a proxy in the
CP and allied models.
Ultimately, the clean CP is bound to enter the absorbing state, the
approach of which is characterized by an exponentially decaying
probability of survival, $P_s(t) \sim \exp(-t/\tau)$, with
a characteristic lifetime $\tau$.
In the active state, $\Delta>0$, this lifetime is known to obey
finite-size scaling,
\begin{eqnarray}
\tau \sim N^{z} \exp \left( N^d \Delta^{\nu_{\perp}d} \right)~,
\label{eq:tau_clean_scaling}
\end{eqnarray}
where $z$ and $\nu_{\perp}$ are critical exponents while $N$ denotes
the size of the system.
Further, the QS density of infected sites, $\overline{\rho}$, obeys a
similar scaling form characterized by the universal scaling function
$F$, i.e. 
\begin{eqnarray}
\overline{\rho} \sim N^{-x} F \left( \Delta
  N^{-1/\nu_{\perp}}\right)~,
\label{eq:rho_fss}
\end{eqnarray}
with $x=\beta / \nu_{\perp}$, where $\beta$ is the order parameter
exponent in the infinite system defined via the behavior of the
control parameter in the vicinity of the critical point, $\rho \sim
\Delta^{\beta}$ \cite{Marro_99:book,dickman_98}.

Turning to the disordered process, we follow Ref.~\cite{Vojta_05} and
incorporate quenched randomness into the transmission rates of
individual sites $i$, $\lambda_i$, which are drawn from the bimodal distribution
\begin{equation}
P(\lambda_i) = (1-p) \delta(\lambda_i-\lambda) + p \delta(\lambda_i-c \lambda)~,
\label{eq:bimodal}
\end{equation}
where $p$ controls the concentration of impurities while $0<c<1$
characterizes the strength of the disorder.
As $c<1$, a particular realization of the disorder contains a
concentration $p$ of randomly arranged impurity sites which 
are less active than the surrounding sea of host sites.
As a consequence, there now exists a new dirty critical point at a
rate $\lambda_c > \lambda_c^0$.
Also, observables may in general take different values between different
realizations leading to disorder-induced distributions such as
$P(\tau)$ for the lifetime.

Scaling predictions for $P(\tau)$ exist and differ for the two
alternative scaling scenarios, i.e. conventional and activated scaling.
In the former case, the relevant variable is $\tau$ and its average
over disorder, $\langle \tau \rangle$, is expected to obey a scaling
form analogous to Eq.~(\ref{eq:tau_clean_scaling}) albeit with
a possibly disorder-dependent dynamical exponent $z$.
Accordingly, the appropriate scale-invariant combination of variables 
is $\tau N^{-z}$ and the lifetime distribution at criticality 
 is expected to scale as
\begin{eqnarray}
P(\tau) = N^{-z} \tilde{P}(\tau N^{-z})~,
\label{eq:tau_conv_scaling}
\end{eqnarray}
where $\tilde{P}$ is a universal scaling function. 

Systems that exhibit activated scaling are characterized by a strong dynamical
anisotropy: the typical length-scale is related to the logarithm of the 
typical timescale, thus rendering the dynamical exponent formally
infinity.
This reflects the notion that the very broad distributions 
for observables are better described by their geometric
rather than arithmetic means \cite{igloi_05}.
For the lifetime, this leads to a scaling combination $N^{-\Psi}\ln(\tau)$ and a
corresponding scaling relation \cite{Hooyberghs_03},
\begin{eqnarray}
P(\ln(\tau)) \sim N^{-\Psi} \tilde{P}(N^{-\Psi} \ln(\tau))~,
\label{eq:tau_act_scaling}
\end{eqnarray}
where $\Psi$ is an activated scaling exponent 
[cf. Eq.~(\ref{eq:tau_conv_scaling})].
For an IRFP, which is known to control the
critical behavior of the DCP for sufficiently strong disorder, the
exponent takes the value $\Psi=\frac{1}{2}$.
This type of fixed point is characterized by an extreme dynamical
anisotropy, ultra-slow dynamics and distributions whose width diverges
with system size \cite{igloi_05,vojta_06_review}.

The QS density $\overline{\rho}$ is expected to follow a conventional
scaling form analogous to Eq.~(\ref{eq:rho_fss}) with an exponent
$x$ which differs from the clean DP value and, 
for sufficiently strong disorder, is predicted to be
$x=\frac{3-\sqrt{5}}{4} \approx 0.19$ \cite{Hooyberghs_03}.

In order to check the above scaling relations for the DCP, the QS
state can be investigated numerically.
Analysis of this metastable state in computer simulations has proved
to be notoriously difficult in the past.
Commonly, the time-dependent density of infected sites conditioned on
survival, which becomes stationary in the QS regime, is investigated
\cite{Marro_99:book}.
Problematically though, it is neither clear a priori at what time an
observable like the average density of infected sites
$\overline{\rho}$ has converged to
its QS value nor when the QS state starts to decay due to
finite size effects \cite{Lubeck_2003}.
Therefore, a range of alternative approaches have been proposed which
enable an observation of this metastable regime (see
Ref.~\cite{oliveira_2005} and references therein).
Here, we employ the QS simulation method \cite{oliveira_2005} which
allows a direct sampling of the QS state by eliminating the absorbing
state and redistributing its probability mass over the active states
according to the history of the process.
The modified process possesses a true stationary state which
corresponds to the original QS state and allows a precise measurement
of QS observables.
Generally, the method has proved to be efficient with fast
and reliable convergence after optimization of history sampling parameters.
As demonstrated in Ref.~\cite{Oliveira2005}, the lifetime of the
QS state can be determined as the inverse of the rate of
attempts by the system to enter the absorbing state. 

\begin{figure}[h!]
\flushleft
\subfigure[$\quad c=0.2$, $\Psi=0.29$ and $z=3.44$]{
\scalebox{0.54}{\includegraphics[scale=1]{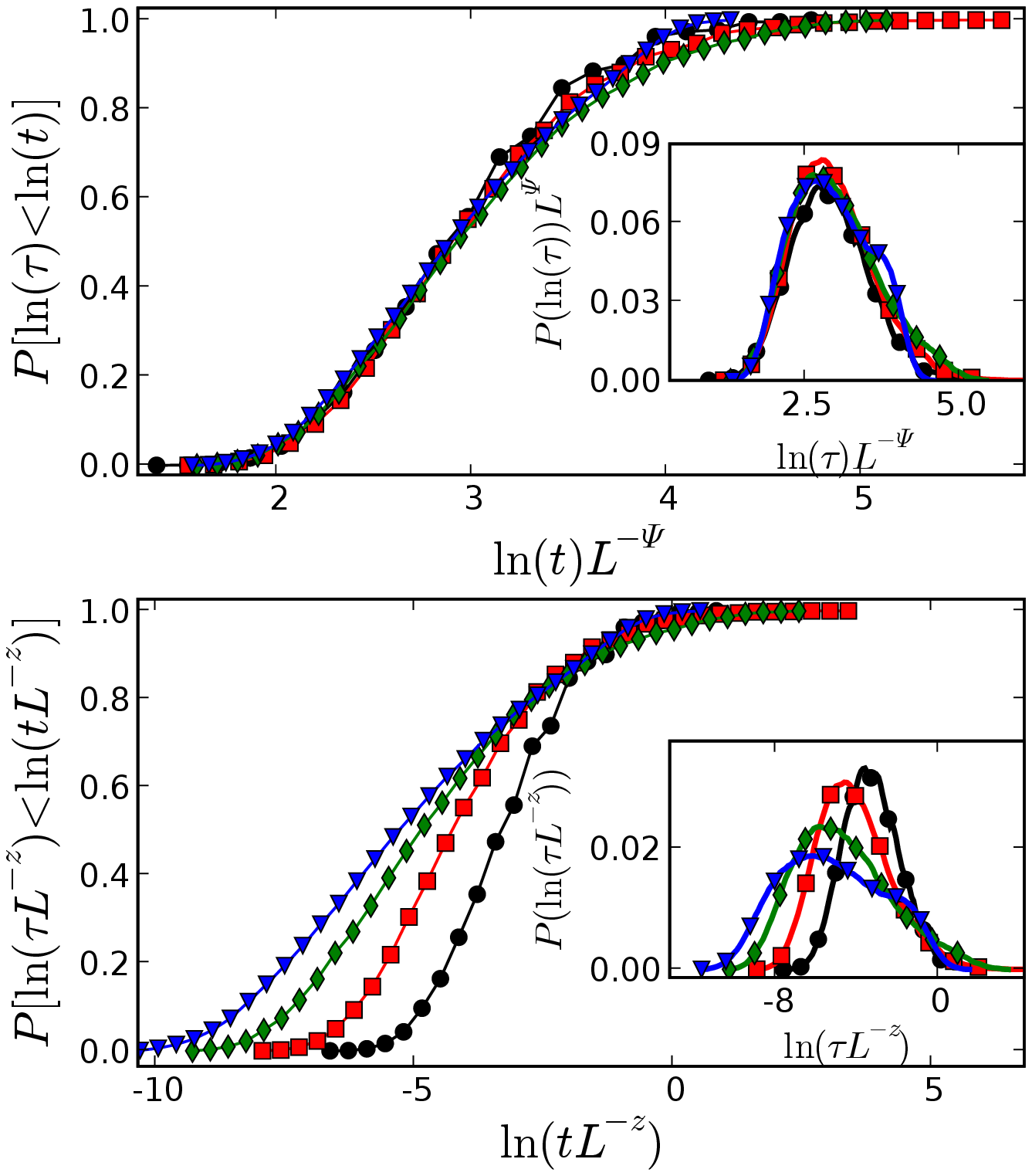}}
\label{fig:tau_dirty_c0.2}
}
\subfigure[$\quad c=0.8$, $\Psi=0.22$ and $z=1.65$]{
\scalebox{0.54}{\includegraphics[scale=1]{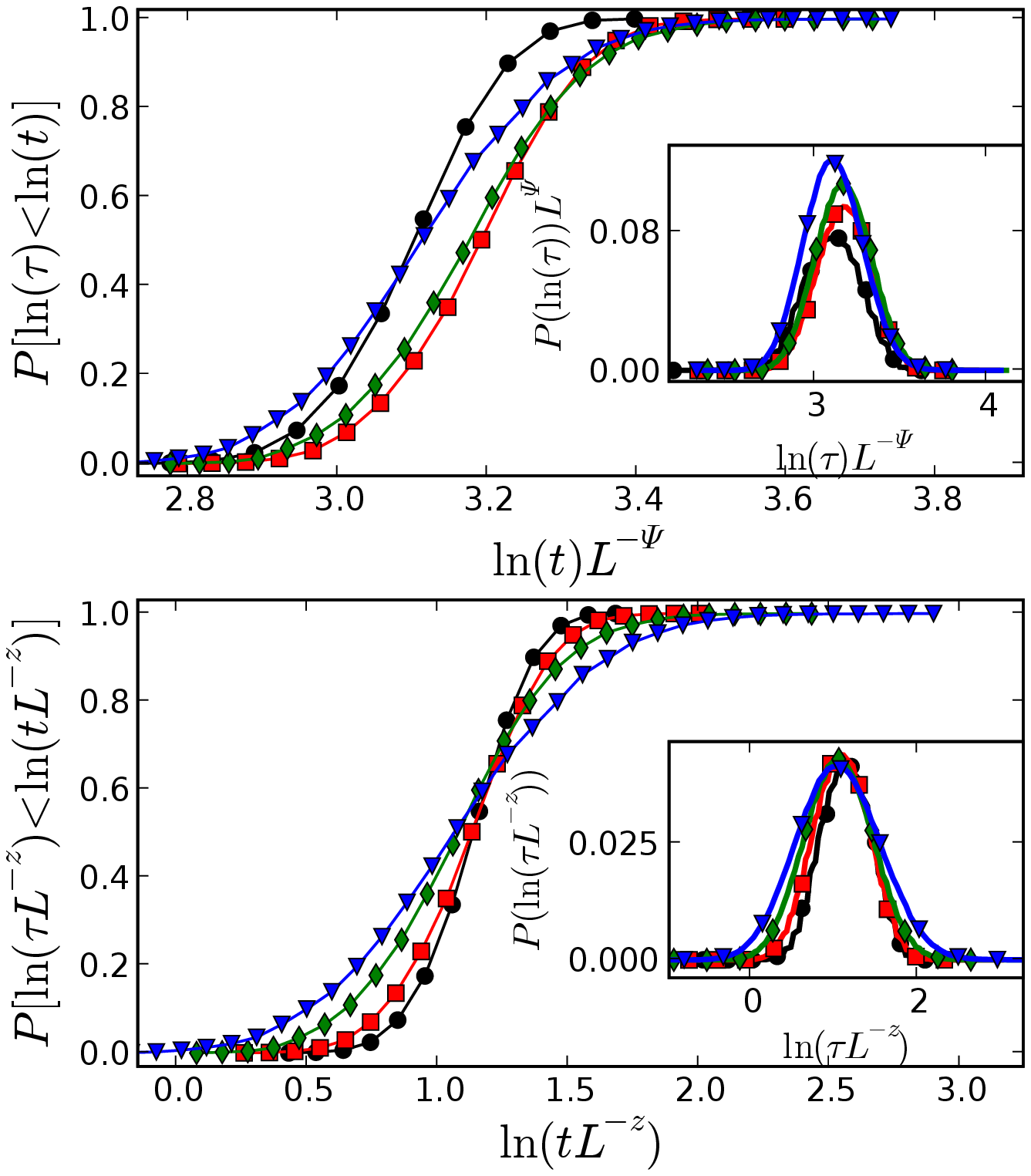}}
\label{fig:tau_dirty_c0.8}
}
\caption[]{(Color online) Scaling collapse for the distributions of
  lifetimes $\tau$ for system sizes $N=16\,(\circ), 32\,(\Box),
  64\,(\Diamond), 128\,(\nabla)$ for disorder strengths $c=0.2$
  (a) and $c=0.8$ (b) according to both activated (main panel) and
  conventional (inset) scaling predictions. Scaling exponents
  were determined from the finite-size scaling of the means $\langle
  \ln(\tau) \rangle$ or $\langle \tau \rangle$, respectively, as
  described in the text.}
\label{fig:tau_dirty}
\end{figure}

In order to investigate the validity of the two different scaling
scenarios for the lifetime distribution at the dirty critical point,
simulation data for disorder strengths $c=0.2, 0.4, 0.6$ and $0.8$ at
a concentration $p=0.3$ were considered at the critical rates reported
in Ref.~\cite{Vojta_05}.
System sizes of $N=16,32,64$ and $128$ sites were investigated with data
sampled from no less than $10^4$ disorder realizations per system
size and QS simulation times of up to $10^9$ time steps.
Given that probability density functions (PDFs) are difficult to
obtain from simulation data (as binning procedures have to be used
which may introduce artifacts), we perform scaling on
cumulative density functions (CDFs), 
$F_{\tau}(t) = \int_0^t P(t')\,\text{d}t'$.
The scaling properties of these can be derived by starting from the
two scaling forms for conventional and
activated scaling, given by Eqs.~(\ref{eq:tau_conv_scaling}) and
(\ref{eq:tau_act_scaling}), respectively, where for the former case
\begin{eqnarray}
F_{\tau}(t) = \int_{0}^{t} N^{-z}~\tilde{P}(t' N^{-z}) \,\text{d} t' 
= \tilde{Q}(tN^{-z})~,
\end{eqnarray}
with $\tilde{Q}(x)$ a new scaling function. 
An analogous expression follows for the case of activated scaling with
$\tau$ replaced by $\ln(\tau)$.

\begin{figure}[t!] %
 \begin{center}
 \scalebox{0.42}{\includegraphics[angle=0]{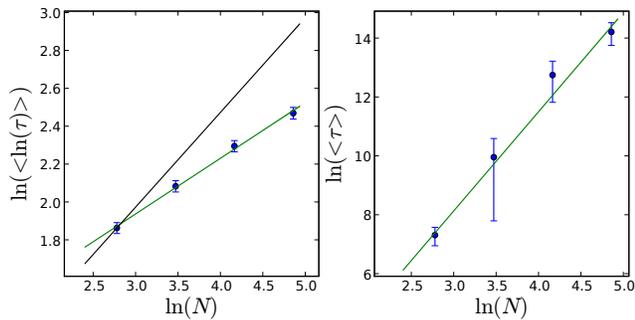}}
 \end{center}
 \caption{(Color online) Example of the scaling of the average
   $\langle \ln(\tau) \rangle$ (left) and $\langle \tau \rangle$
   (right) with system size for the case of strongest disorder
   ($c=0.2$) used to extract the exponents $\Psi$ and $z$.}
 \label{fig:tau_exponents}
 \end{figure}

As displayed in Fig.~\ref{fig:tau_dirty}, the resulting CDFs were
collapsed onto each other according to the two possible scaling
scenarios (main panels).
In order to achieve a fair comparison, logarithmic variables were used
for the conventional scaling case \cite{Young_96}.
As illustrated in Fig.~\ref{fig:tau_exponents}, the exponents $\Psi$
  and $z$ were determined from a power-law fit to the
appropriate scaling forms for the mean in the two scenarios,
i.e. $\langle \ln(\tau) \rangle \sim N^{\Psi}$ and $\langle \tau
\rangle \sim N^{z}$.
Insets show the alternative collapse using PDFs as discussed above
which requires the use of histograms.
There,  the size of bins was chosen in order
to minimize noise and the smooth curve was obtained by Gaussian
broadening of individual data points.

For the case of strongest disorder ($c=0.2$, $\lambda_c=5.24$),
least-squares fitting gave
exponents $\Psi_{c=0.2}=0.29(2)$ and $z_{c=0.2}=3.44(3)$ for the two scaling
scenarios, respectively.
Data collapses for the distributions are shown in
Fig.~\ref{fig:tau_dirty_c0.2} for both activated (top panel) and
conventional scaling (bottom panel).
From these results, we judge that the activated scaling scenario
provides a better fit to the data.
In particular, while the collapse is not perfect, it is not found to
exhibit any systematic trends which would hint at a fundamental
inconsistency with the scaling form.
This is also confirmed by the inset which shows the corresponding
collapse of the PDF.
Generally, while the fit is excellent for small to medium values of
$\tau$, it gets worse with increasing $\tau$.
We attribute this to the fact that large values of $\tau$
correspond to rare events causing the tail of the distribution
to have been sampled at a comparatively poorer density than the bulk
which results in stronger fluctuations.
Considering the collapse using the conventional scaling form [bottom
panel of Fig.~\ref{fig:tau_dirty_c0.2}] on the
other hand produces a clear trend of shifting of distributions between
different system sizes indicating a worse collapse as compared to the
previous case for both CDF and PDF.

Looking at an analogous analysis for the case of weakest disorder
($c=0.8$, $\lambda_c=3.525$) in Fig.~\ref{fig:tau_dirty_c0.8},
the two scaling scenarios become harder to differentiate.
A collapse using the measured exponents
$\Psi_{c=0.8}=0.22(2)$ and $z_{c=0.8}=1.65(3)$ appears to work
similarly well in both cases but the quality of collapse is too poor
to allow any definitive judgment.
A close look reveals a systematic trend of shifting curves
for the case of conventional scaling [bottom panel in
Fig.~\ref{fig:tau_dirty_c0.8}] while crossings appear to be
less systematic in the activated scaling picture [top panel in
Fig.~\ref{fig:tau_dirty_c0.8}].
Therefore, slight preference may be given to the activated
scaling scenario.

In addition, the intermediate cases of both $c=0.4$ and $c=0.6$
(not shown) were considered in an analogous fashion and showed an excellent
activated scaling collapse of similar quality as for the case of $c=0.2$.
Collapsing data for all disorder strengths considered on the
same plot shows no universality of the scaling function between
different disorder strengths, i.e. it appears to be disorder dependent.

 \begin{figure}[t!] %
 \begin{center}
 \scalebox{0.45}{\includegraphics[angle=0]{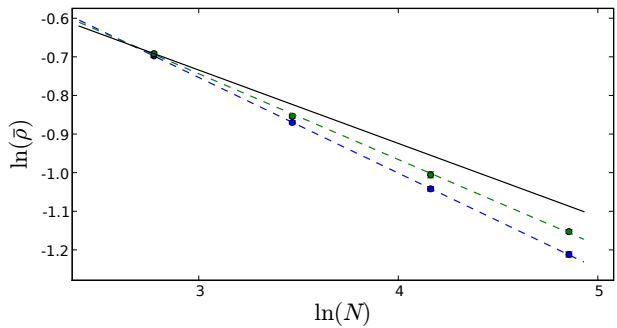}}
 \end{center}
 \caption{(Color online) The quasi-stationary density $\overline{\rho}$
   as a function of system size $N$ at criticality for both the
   strongest (top, $c=0.2$) and the weakest (bottom: $c=0.8$) disorder
   strength. Dashed lines are linear least-squares
   fits which gave slopes of $x_{c=0.2}=0.22(1)$ and
   $x_{c=0.8}=0.25(1)$ while the solid line is a guide to the eye with 
   slope
   $x_{\text{strong}}=0.19$, the expected exponent in the strong-disorder
   limit.}
 \label{fig:rho_fss}
 \end{figure}

\begin{figure}[t!] %
\begin{center}
\scalebox{0.4}{\includegraphics[angle=0]{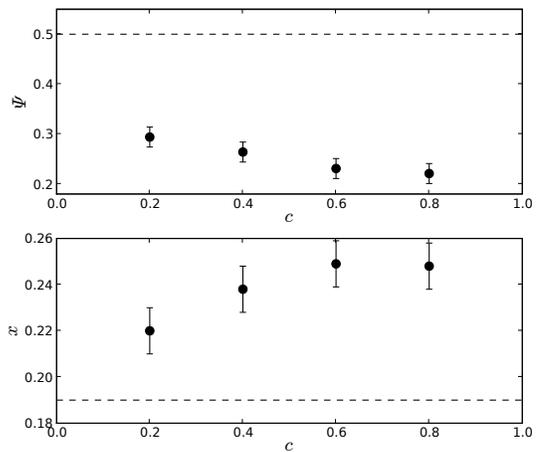}}
\end{center}
\caption{(Color online) Critical exponents $\Psi$ (top panel) and $x$
  (bottom panel) as a function of disorder strength $c$ where dashed
  lines show the values in the limit of strong disorder.}
\label{fig:exponents}
\end{figure}

Finite-size scaling of the QS density $\overline{\rho}(N)$, as shown
in Fig.~\ref{fig:rho_fss}, is found to be conventional with a
disorder-dependent exponent $x=x(c)$.
The conventional scaling form is in line with previous investigations 
\cite{dickman_98} and theoretical predictions \cite{Hooyberghs_03}.
Moreover, the fact that we again find continuously varying exponents gives
additional credibility to the scaling picture presented above.
Generally, both the exponent $\Psi$ and the exponent $x$ are found to
approach their values predicted from strong-disorder renormalization
\cite{Hooyberghs_03} with increasing strength of disorder starting
from their homogeneous values as shown in Fig.~\ref{fig:exponents}.
For the strongest disorder under consideration ($c=0.2$), both
exponents are found to still be well away from their predicted
strong-disorder values.

The above findings suggest that for strong enough disorder, activated
scaling captures the behavior of the disordered CP well compared to
a conventional scaling picture.
For weak disorder however, no such clear conclusion can be made.
Further, associated critical exponents appear to change smoothly from
their clean DP values approaching their values characteristic of an
IRFP asymptotically with increasing disorder strength.
While this conclusion appears to be in conflict to that presented in
Ref.~\cite{Vojta_05}, the authors of that reference do discuss doubts
about the universality of exponents and cannot exclude the possibility
of a change with disorder strength.
We have re-analyzed some of the data presented there and found it to
be compatible with our exponent values.

There exist three possible explanations compatible with these
findings. 
First, a continuous line of fixed points, one for each
strength of disorder, could be present which for sufficiently strong
disorder turns to an attractive flow into the IRFP as suggested in
Ref.~\cite{Hooyberghs_03}.
Second, identical and numerically indistinguishable behavior could be
explained by a crossover between the clean DP fixed point and the IRFP
where effective exponents are observed at intermediate disorder
strengths due to the influence of both fixed points.
This has been observed in several disordered equilibrium systems as
discussed in e.g.\ Refs.~\cite{Carlon_01,Fisher_74}.
Lastly, in principle, the observed behavior could also be explained by
an abrupt jump from the clean DP exponents to those of the IRFP
obscured by finite-size corrections.

The last option we feel can be excluded in light of the facts that
perturbative series expansions (cf.\ Ref.~\cite{Neugebauer_06}) do not
show a jump in exponents and that no evidence for strong corrections
to finite-size scaling were observed by us.
At the same time, the other two scenarios are compatible with our and most
other results but cannot be safely distinguished by
numerical investigation alone without an established theoretical
framework for the crossover in the DCP.

We would like to thank Ronald Dickman for
helpful remarks. Also, we thank Allon Klein, Chris Neugebauer and
Francisco P\'{e}rez-Reche for stimulating discussions.
The computations were performed on the Cambridge High Performance
Computing Facility. SVF would like to thank the UK EPSRC and the
Cambridge European Trust for financial support.

\bibliographystyle{apsrev}


\end{document}